\documentclass[showpacs,twocolumn]{revtex4}
\usepackage{amsmath,amssymb}
\usepackage{natbib}
\usepackage{url}
\usepackage{graphicx}
\usepackage{psfrag}
\usepackage{color}
\usepackage{epstopdf}

\usepackage{mathrsfs}
\DeclareSymbolFontAlphabet{\mathrsfs}{rsfs}
\DeclareMathAlphabet{\mathcal}{OMS}{cmsy}{m}{n}

\newcommand{\scri}{\mathrsfs{I}}
\newcommand{\be}{\begin{equation}}
\newcommand{\ee}{\end{equation}}

\def\rhoo{\rho_{\mathrm{e}}}
\def\rhoi{\rho_{\mathrm{i}}}
\def\Hext{H_{\mathrm{e}}}
\def\Hint{H_{\mathrm{i}}}

\def\balpha{\bar{\alpha}}
\def\bbeta{\bar{\beta}}
\def\bJ{\bar{J}}
\def\bc{\bar{c}}

\begin{document}
\title{Asymptotics of Schwarzschild black hole perturbations}

\author{An{\i}l Zengino\u{g}lu} 

\affiliation{Department of Physics, Center for Fundamental Physics, \\
 and Center for Scientific Computation and Mathematical Modeling, \\
 University of Maryland, College Park, MD 20742, USA}

\email{anil@umd.edu}

\begin{abstract}
We study linear gravitational perturbations of Schwarzschild spacetime by solving numerically Regge-Wheeler-Zerilli equations in time domain using hyperboloidal surfaces and a compactifying radial coordinate. We stress the importance of including the asymptotic region in the computational domain in studies of gravitational radiation. The hyperboloidal approach should be helpful in a wide range of applications employing black hole perturbation theory.
\end{abstract}

\maketitle

\section{Introduction}
Linear perturbation theory of black holes is a well developed theory \cite{Chandrasekhar:1992bo, Kokkotas:1999bd, Nagar:2005ea}. The most common approach to linear perturbations of a non-rotating single black hole is to solve the Regge-Wheeler-Zerilli (RWZ) equations for a master function built from a tensor spherical harmonic decomposition of metric perturbations \cite{Regge:1957rw, Zerilli70}. The RWZ formalism plays a prominent role in many studies ranging from linear stability of black holes \cite{Regge:1957rw, Zerilli70, Vishveshwara70} to extreme mass ratio inspirals \cite{Martel:2003jj, Sopuerta:2005rd, Jung:2007zf, Field:2009kk, Canizares:2009ay, Blanchet:2009sd}. 

An important problem in numerical time domain calculations of black hole perturbations is the truncation of the infinite spatial domain of Schwarzschild time slices to a finite domain by the introduction of a timelike, artificial outer boundary. The presence of such a boundary implies an initial boundary value problem. As the boundary is not part of the physical solution one tries to construct transparency boundary conditions so that the numerical solution is as close as possible to the physical solution without the boundary. 

The construction of transparent boundary conditions is a difficult problem, especially in general relativity  \cite{Friedrich:1998xt, Friedrich:2009tq, Rinne:2007ui,Seiler:2008hm, Nunez:2009wn}. The RWZ formalism plays a central role in certain approaches to this problem, for example in Cauchy-perturbative matching \cite{Rupright98, Rezzolla99a, Zink:2005pe}, or the implementation of Buchman-Sarbach boundary conditions  \cite{Buchman:2006xf, Buchman:2007pj, Rinne:2008vn}.

It is not straightforward, however, to construct good boundary conditions even for the RWZ equations. A well developed method to deal with this difficulty has been presented by Lau \cite{Lau:2004as, Lau:2004jn, Lau:2005ti}. Lau constructs exact radiation outer boundary conditions for RWZ equations based on a study of time-domain boundary kernels in Schwarzschild spacetime (see also the underlying approach developed by Alpert, Greengard and Hagstrom in flat spacetime \cite{Alpert00}). Lau's method makes it possible to perform accurate, long-time simulations of RWZ equations in a small spatial domain, thereby increasing the efficiency of numerical calculations significantly. 

On the other hand, truncating the computational domain for studying gravitational radiation complicates the access to asymptotics of black hole perturbations considerably. In studies of gravitational perturbations the main interest lies in the radiation signal as observed at future null infinity. In simulations on spatially compact domains, elaborate methods are applied on timelike surfaces that represent far away observers to extract a signal that is as close to the true signal at null infinity as possible \cite{Deadman:2009ds, Gallo:2008sk, Nerozzi:2008ng, Barton:2008eb, Boyle:2009vi}.  But even when the background is given and gauge independent extraction methods can be conceived, it is important to have access to the asymptotic domain to capture the nonlinear self interaction of the field away from the sources, as argued in Sec.~\ref{sec:decay}.

Including null infinity in the computational domain allows us to calculate the asymptotics of gravitational fields and deems unnecessary the problem of constructing radiation outer boundary conditions. Only recently, the first unambiguous waveform from merging black holes could be numerically calculated along null infinity using a mixture of Cauchy and characteristic
methods \cite{Reisswig:2009us, Reisswig:2009vc}. An alternative,
efficient approach to include null infinity in the computational
domain is to compactify spacelike surfaces that approach future null
infinity \cite{Friedrich83a, Huebner01, Husa01, Frauendiener04, Friedrich04}. Such surfaces are called hyperboloidal as their asymptotic behavior resembles the asymptotic behavior of standard hyperboloids in Minkowski spacetime. There has been growing interest into the study of hyperboloidal foliations
\cite{Malec:2003dq, Malec:2009hg, Zenginoglu:2007jw, Ohme:2009gn, Buchman:2009ew} and numerical solutions to the hyperboloidal initial value problem on given backgrounds  \cite{Fodor:2003yg,Fodor:2006ue, vanMeter:2006mv, Bizon:2008zd, Zenginoglu:2008wc,Zenginoglu:2008uc,Gonzalez:2009hn,Zenginoglu:2009hd}.

The hyperboloidal method has been shown to be very efficient in studies of gravitational perturbations \cite{Zenginoglu:2008uc}. The authors of \cite{Zenginoglu:2008uc} solve the Bardeen-Press equations based on the Newman-Penrose formalism \cite{Bardeen73a, Newman62a}. In this paper, we extend the hyperboloidal study of gravitational perturbations to include the RWZ formalism. A basic motivation for this extension is that the RWZ formalism is a more common tool to study gravitational perturbations of a single black hole due to its simplicity.

The extension of the hyperboloidal method to the RWZ formalism allows us to compare tail decay rates of gravitational perturbations within the Bardeen-Press and the RWZ formalisms. The comparison reveals insights into the validity of asymptotic formulae on gravitational radiation and demonstrates the importance of including the asymptotic domain in numerical calculations.

We extend the study \cite{Zenginoglu:2008uc} also with respect to the hyperboloidal method such that null infinity can be fixed at an arbitrary radial coordinate location, and arbitrary coordinates can be used near the black hole \cite{Zenginoglu:2007jw, Zenginoglu:2009hd}. We find that the application of a transition function suggested in \cite{Yunes:2005nn, Vega:2009qb} gives a more accurate spacelike matching method than the one suggested in \cite{Zenginoglu:2007jw, Zenginoglu:2009hd}. These improvements are essential for dealing with matter terms near the black hole within the hyperboloidal approach.

\section{Schwarzschild spacetime in scri-fixing coordinates}
\label{sec:2}

The following section is based on \cite{Zenginoglu:2007jw} but evades the conformal language used in that reference. We present the three steps of the hyperboloidal scri-fixing compactification in Schwarzschild spacetime: choice of a hyperboloidal time function, introduction of a compactifying coordinate, and rescaling of metric functions with a suitable conformal factor.
\subsection{Choice of foliation}
\begin{figure}[t]
  \centering
  \psfrag{sing}{\footnotesize{singularity}} \psfrag{hor}{$\mathcal{H}^+$}
  \psfrag{ip}{$i^+$} \psfrag{im}{$i^-$} \psfrag{i0}{$i^0$}
  \psfrag{scrp}{$\scri^+$} \psfrag{scrm}{$\scri^-$}\psfrag{phor}{$\mathcal{H}^-$}
  \includegraphics[height=0.23\textheight]{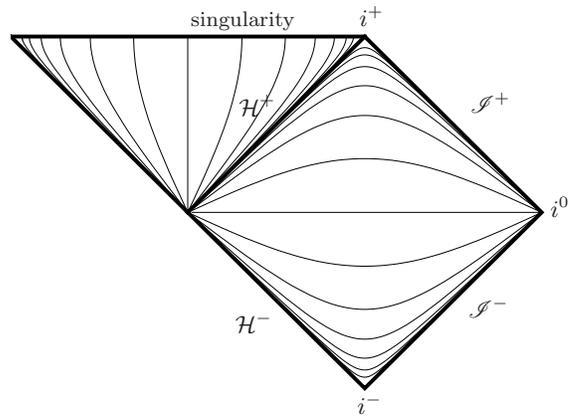}
  \caption{Level surfaces of the standard Schwarzschild time
    coordinate meet at the bifurcation sphere and at spatial infinity. \label{fig:ss}}
\end{figure}

The Schwarzschild metric in standard coordinates $(t,r)$ reads
\[ g=-F\,dt^2 + F^{-1}\,dr^2+r^2\,d\sigma^2,  \quad \textrm{with}\quad 
F:=1-\frac{2m}{r}. \] Here, $d\sigma^2$ is the standard metric on the unit sphere. The mass of the Schwarzschild black hole is denoted by $m$. Hypersurfaces of constant time coordinate all meet at the bifurcation sphere near the black hole and at spatial infinity in the asymptotic domain (Fig.~\ref{fig:ss}). The coordinates are singular at the event horizon $r=2m$. There are two common ways to deal with this problem. One either introduces the tortoise coordinate and applies outgoing radiation boundary conditions close to the black hole, or one switches to horizon penetrating coordinates and applies an excision technique. 

It is instructive to discuss horizon penetrating coordinates and to
compare them with hyperboloidal coordinates. For both gauge classes,
we change the foliation by introducing a new time function
\be\label{time_traf} \tau = t-h(r). \ee All foliations discussed in
this paper are of the above form. The new foliation respects the
stationarity of Schwarzschild spacetime. The transformed metric can be written
as \be \label{eq:decomp} g =(-\alpha^2+\gamma^2\beta^2)\,d\tau^2 + 2
\gamma^2 \beta\, d\tau\,dr +\gamma^2\,dr^2+r^2\,d\sigma^2,  \ee where the lapse
$\alpha$, the shift $\beta$, and the spatial metric function $\gamma$
read \be\label{tr_gauge} \alpha^2 = \frac{F}{1-(F H)^2},\quad \beta =
- F H \,\alpha^2, \quad \gamma = \frac{1}{\alpha}, \ee with
$H:=\frac{dh}{dr}$ called the height function derivative.

\begin{figure}[t]
  \centering \psfrag{sing}{\footnotesize{singularity}}
  \psfrag{hor}{$\mathcal{H}^+$}\psfrag{phor}{$\mathcal{H}^-$}
  \psfrag{ip}{$i^+$} \psfrag{im}{$i^-$} \psfrag{i0}{$i^0$}
  \psfrag{scrp}{$\scri^+$} \psfrag{scrm}{$\scri^-$}
  \includegraphics[height=0.23\textheight]{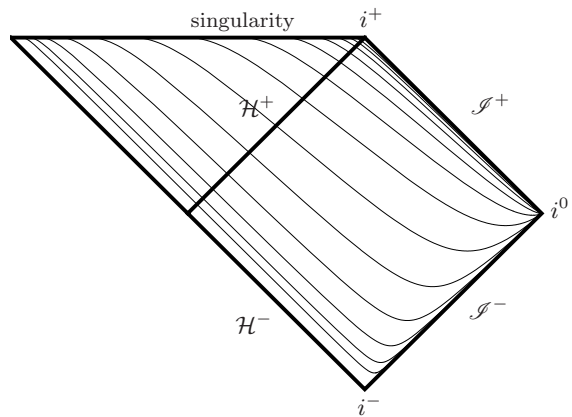}
  \caption{Level surfaces of the ingoing Eddington-Finkelstein time
    coordinate (\ref{height_ef}) foliate the future
    event horizon, but meet at spatial infinity in the asymptotic domain. \label{fig:ief}}
\end{figure}

A common system of horizon penetrating coordinates in
Schwarzschild spacetime is based on ingoing Eddington-Finkelstein
surfaces \cite{Eddington24, Finkelstein58}. The height function
derivative for an ingoing Eddington-Finkelstein foliation reads
\be\label{height_ef} H_\textrm{iEF} = -\frac{2m}{r-2m}. \ee The time
surfaces foliate the future event horizon (Fig.~\ref{fig:ief}) and all
metric components are regular at \mbox{$r=2m$}.

The construction of hyperboloidal coordinates is analogous to the
construction of horizon penetrating coordinates. Comparing
Figs.~\ref{fig:ss} and \ref{fig:ief} we see that a height function
with a suitable singular behavior at the horizon (\ref{height_ef})
opens the time surfaces at the bifurcation sphere so they foliate the
future event horizon. Similarly, a height function with a suitable
singular asymptotic behavior opens the time surfaces at spatial infinity so they foliate future null infinity. Expanding the asymptotic hyperboloidal condition in a Taylor series gives
\cite{Zenginoglu:2007jw} \be\label{height_taylor} H_\textrm{Taylor} =
1+\frac{2m}{r} + \left(4m^2-\frac{C_T^2}{2}\right)\frac{1}{r^2} +
O\left(\frac{1}{r^3}\right).\ee Here, $C_T$ is a free constant. The explicit form of the condition depends naturally on
the coordinates in which it is expressed. We see that the height function to (\ref{height_taylor}) diverges in the asymptotic domain.

The first discussion of hyperboloidal foliations in general relativity
has been made for surfaces with constant mean curvature (CMC)
\cite{York79,Eardley79}. Spherically symmetric CMC surfaces have been
constructed explicitly in Schwarzschild spacetime by Brill, Cavalho
and Isenberg \cite{Brill80} and studied in much detail by Malec and
O'Murchadha \cite{Malec:2003dq,Malec:2009hg}. The height function
derivative for such surfaces reads \be\label{height_cmc}
H_\textrm{CMC}=\frac{J} {F \sqrt{J^2+F}}, \quad \textrm{where} \quad J
:= \frac{K r}{3}-\frac{C}{r^2}.\ee The foliation parameters are the
mean extrinsic curvature $K$ and a constant of integration $C$. We are
not interested in varying $K$ or $C$ from one slice to another
\cite{Malec:2003dq,Malec:2009hg}. Instead, we drag the hypersurfaces
along the timelike Killing vector by keeping the parameters fixed.

\begin{figure}[t]
  \centering 
  \psfrag{sing}{\footnotesize{singularity}}\psfrag{phor}{$\mathcal{H}^-$}
  \psfrag{hor}{$\mathcal{H}^+$} \psfrag{ip}{$i^+$} \psfrag{im}{$i^-$}
  \psfrag{i0}{$i^0$} \psfrag{scrp}{$\scri^+$} \psfrag{scrm}{$\scri^-$}
  \includegraphics[height=0.23\textheight]{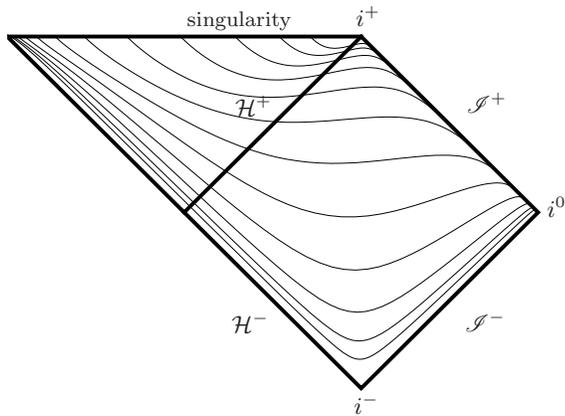}\hspace{1cm}
  \caption{CMC-foliation of the extended Schwarzschild spacetime with
  $K=0.4$, $C=2$ and $m=1$. The time surfaces foliate the future event horizon as well as the future null infinity. \label{fig:cmc}}
\end{figure}

The global behavior of CMC-surfaces depends crucially on the foliation
parameters. For numerical applications, we would like to have surfaces that approach future null infinity asymptotically (for radiation extraction) and penetrate the future horizon near the black hole (for excision). These requirements can be fulfilled by choosing $K>0$ and $C>8m^3K/3$ \cite{Malec:2003dq, Buchman:2009ew}.  Such a slicing of Schwarzschild spacetime is depicted in Fig.~\ref{fig:cmc}.

The asymptotic behaviour of the height function derivative for CMC surfaces (\ref{height_cmc}) with $K>0$ is given by
\[ H_\textrm{CMC} = 1+\frac{2m}{r}+\left(4m^2-\frac{9}{2K^2}\right)\frac{1}{r^2}
+ O\left(\frac{1}{r^3}\right).\] Comparing this with
(\ref{height_taylor}) we see that the parameter $C_{T}$ is related to
the mean extrinsic curvature in lowest order by $C_T=3/K$.

The height function derivative for ingoing Eddington-Finkelstein coordinates, $H_\textrm{iEF}$ in (\ref{height_ef}), vanishes
asymptotically because ingoing Eddington-Finkelstein surfaces approach
spatial infinity. This is also true for maximal surfaces with
$K=0$, which suggests that the mean extrinsic curvature can
be regarded, in a sense, as a measure of how close the CMC
surfaces are to being Cauchy or characteristic
\cite{Zenginoglu:2007jw}. It also controls the strength of the spatial redshift effect discussed in \ref{sec:match}.

Inserting the height function derivative (\ref{height_cmc}) into
(\ref{tr_gauge}) gives
\be\label{cmc_metr} \alpha = \sqrt{J^2+F},\quad \beta = -J \alpha.\ee
We observe an important difference between Cauchy and hyperboloidal
foliations.  The asymptotic behavior of an asymptotically flat
spacetime written in a Cauchy foliation implies $\alpha \sim 1$ and $\beta \sim 0$ as $r\to \infty$. The same spacetime in a hyperboloidal foliation satisfies
\cite{York79} \be\label{eq:asympt} \alpha \sim O(r) \quad \mathrm{and}
\quad \beta \sim O(r^2) \quad \mathrm{as} \quad r\to\infty.\ee We need
to take this singular behavior into account when introducing a
compactifying radial coordinate in the next subsection.

\subsection{Compactification and rescaling}
To map future null infinity to a finite coordinate location, we choose
a compactifying coordinate along future hyperboloidal surfaces. We define the compactifying coordinate $\rho$ by \be\label{eq:rcomp} r= \frac{\rho}{\Omega}, \quad \textrm{with}
\quad \Omega=\Omega(\rho),\ee so that the set $\{\Omega(\rho)=0\}$
corresponds to infinity in terms of the coordinate $r$. The above transformation is written in a generality that allows us to use the
standard coordinate $r$ in any given domain $\mathcal{D}$ by setting
$\Omega(\rho)=1$ for $\rho\in\mathcal{D}$. 

We choose $\Omega$ such that asymptotically $\Omega\sim 1/r$. We introduce the rescaling \cite{Zenginoglu:2008uc} 
\be\label{def_gauge}\balpha := \Omega
\,\alpha, \qquad \bbeta := \Omega^2\beta.\ee The rescaled lapse and shift are not, in general, the lapse and shift of the conformally rescaled Schwarzschild spacetime. The above rescaling is just a mathematical trick to replace the singular lapse and shift by functions that are regular at null infinity. For example, for the CMC foliation with the height function (\ref{height_cmc}) we get with (\ref{def_gauge})
\be \label{eq:resc} \balpha = \sqrt{\bJ^2+ F\,\Omega^2},\qquad\bbeta
=-\bJ \,\balpha,\ee where
\[ \bJ := \Omega\, J = \frac{K\rho}{3} - \frac{C\,\Omega^3}{\rho^2}.\] 

The regular metric functions $\balpha, \bbeta$ and the rescaling factor $\Omega$ are used in the next section to show the regularity of the RWZ equations in hyperboloidal scri fixing coordinates.

\section{Hyperboloidal compactification for the RWZ equations}\label{sec:3}
The original RWZ formalism assumes a Schwarzschild background in standard Schwarzschild coordinates. We want to solve the RWZ equations with respect to a new time coordinate (\ref{time_traf}), therefore we need a covariant version of the RWZ equations. A gauge invariant formulation of metric perturbations was given by Moncrief \cite{Moncrief74}. Moncrief's formulation also relies on Schwarzschild coordinates. Gerlach and Sengupta presented gauge invariant odd-parity RWZ equations in an arbitrary spherically symmetric coordinate system \cite{Gerlach79, Gerlach:1980tx}. Their formalism has been extended by Sarbach and Tiglio \cite{Sarbach:2001qq}. Martel and Poisson completed the covariant, gauge invariant extension of the RWZ formalism by presenting covariant source terms derived from the stress-energy tensor of the matter perturbations \cite{Martel:2005irw}. We follow the formalism developed by Sarbach and Tiglio \cite{Sarbach:2001qq}. The RWZ equations for a master function $Z$ reads
\be \label{orig_rwz} \partial_t Z_t = c_1 \partial_r Z_t + c_2
\partial_r Z_r + c_3 Z_t + c_4 Z_r - \alpha^2 V Z, \ee where an index on $Z$ denotes partial differentiation with respect to the corresponding coordinate. The coefficients read
\begin{eqnarray} \label{rwz_coef}
c_1 &=& 2\beta, \nonumber \\
c_2 &=& \frac{\alpha^2 - \gamma^2 \beta^2}{\gamma^2}, \nonumber \\
c_3 &=& \frac{1}{\gamma \alpha} (\gamma\partial_t \alpha - \gamma \beta 
\partial_r\alpha+\alpha\beta\partial_r\gamma-\alpha\partial_t\gamma+
\gamma\alpha\partial_r\beta),\nonumber \\
c_4 &=& \frac{1}{\gamma^3 \alpha} (-\gamma^3 \beta \partial_t\alpha-
\alpha^3\partial_t\gamma+\gamma^3\beta^2\partial_r\alpha- 
2\gamma^3\alpha\beta\partial_r\beta+ \nonumber\\
&&\gamma^3\alpha\partial_t\beta+ \gamma^2\alpha\beta\partial_r\alpha-
\gamma^2\alpha\beta\partial_r\gamma).
\end{eqnarray}
The potential $V$ depends on whether we solve for odd (o) or even (e) parity perturbations. The corresponding potentials are given by
\[ V^\textrm{(o)} = \frac{1}{r^2} \left(l(l+1)-\frac{6m}{r}\right),\]
\[ V^\textrm{(e)} = \frac{\lambda^2 r^2 ((\lambda+2)r + 6m) +36 m^2 (\lambda r +2m)}
{(\lambda r + 6m)^2 r^3},\] where $\lambda=(l-1)(l+2)$. Here, $l$ refers to the angular momentum number from the tensor spherical harmonic decomposition of the metric perturbations from which the master function $Z$ is built \cite{Sarbach:2001qq}. Note that both potentials fall off as $1/r^2$ asymptotically.

In this formalism, metric functions may depend both on $t$ and $r$. We do not need the RWZ formalism in such generality. The transformation (\ref{time_traf}) has two properties that simplify the coefficients  (\ref{rwz_coef}) considerably. First, it respects the stationarity of Schwarzschild spacetime. This implies that the metric functions are independent of $\tau$. Second, it leaves the relation $\alpha \gamma = 1$ invariant. Using these properties we get with respect to the new time function $\tau$
\[ \partial_\tau Z_\tau = c_1 \partial_r Z_\tau + c_2 \partial_r Z_r + c_3 Z_\tau 
+ c_4 Z_r - \alpha^2 V Z, \] with coefficients
\begin{eqnarray}
c_1 &=& 2\beta, \nonumber\\
c_2 &=& \alpha^2 \left(\alpha^2 - \frac{\beta^2}{\alpha^2}\right),\nonumber \\
c_3 &=& \alpha^2 \partial_r \left(\frac{\beta}{\alpha^2}\right), \nonumber\\
c_4 &=& \alpha^2 \partial_r \left(\alpha^2 - \frac{\beta^2}{\alpha^2}\right).
\end{eqnarray}

The next step is to introduce the compactifying coordinate $\rho$ given in (\ref{eq:rcomp}) and the rescaled metric functions (\ref{def_gauge}). The resulting RWZ equation reads
\begin{eqnarray} \label{comp_rwz}
\partial_\tau Z_\tau&=&\frac{\bc_1}{L} \partial_\rho Z_\tau + \frac{\bc_2}{L^2}
\partial_\rho Z_\rho + \frac{\bc_3}{L} Z_\tau + \nonumber \\
&& \frac{1}{L^2}\left(\bc_4-\bc_2 \frac{\partial_\rho L}{L}\right) Z_\rho - 
\balpha^2 \bar{V} Z,
\end{eqnarray}
with
\begin{eqnarray}\label{comp_coef}
\bc_1 &=& 2\bbeta, \nonumber\\
\bc_2 &=& \balpha^2 \left(\balpha^2 - \frac{\bbeta^2}{\balpha^2}\right),\nonumber\\
\bc_3 &=& \balpha^2 \partial_\rho \left(\frac{\bbeta}{\balpha^2}\right),\nonumber\\
\bc_4 &=&
\balpha^2\partial_\rho\left(\balpha^2-\frac{\bbeta^2}{\balpha^2}\right),\nonumber\\
\bar{V} &=&\frac{V}{\Omega^2}, \ \qquad\ L= \Omega-\rho\,\partial_\rho \Omega.
\end{eqnarray}
The barred coefficients are related to the original coefficients via
\[ c_1 = \frac{\bc_1}{\Omega^2}, \ c_2 = \frac{\bc_2}{\Omega^4}, \ 
c_3 = \frac{\bc_3}{L}, \ c_4 = \frac{1}{L\Omega^2}\left(\bc_4-\frac{2\partial_\rho 
\Omega} {\Omega}\bc_2\right).\]
The RWZ potentials are regular due to their asymptotic behavior. They read
\be \label{eq:odd_pot} \bar{V}^\textrm{(o)}= \frac{1}{\rho^2}\left(l(l+1)-\frac{6 m\Omega}{\rho}\right),\ee
\[ \bar{V}^\textrm{(e)}=\frac{\lambda^2 \rho^2 ((\lambda+2)\rho+6m\Omega)+36m^2\Omega^2(\lambda\rho+2m\Omega)}{(\lambda\rho+6m\Omega)^2\rho^3}.\]

This ends our discussion of hyperboloidal compactification for the RWZ formalism. The regularity of the system (\ref{comp_rwz}) up to and including null infinity is clear from the regularity of its coefficients (\ref{comp_coef}).

\section{Numerics}\label{sec:4}
\subsection{The computational framework}
The numerical discretization to solve the hyperboloidal initial value
problem for gravitational perturbations (\ref{comp_rwz}) is similar to
what has been used in \cite{Zenginoglu:2008uc}. We use the method of
lines with 4th order Runge-Kutta time integration and 8th order
spatial finite differencing. We add Kreiss-Oliger type artificial
dissipation to the evolution equation for $Z_\tau$ to suppress
numerical high-frequency noise \cite{Kreiss73}. The convergence properties of the code are qualitatively the same as in \cite{Zenginoglu:2008uc}. 

We take a simple Gaussian as the initial perturbation and set
\be\label{eq:id} Z(0,\rho) = 0, \quad Z_\rho(0,\rho)=0, \quad Z_\tau(0,\rho) =
e^{-(\rho-\rho_c)^2/\sigma^2}. \ee Here, $\rho_c$ is the center of the Gaussian
pulse and $\sigma$ is its width. Experiments with different families
of initial data with similarly fast fall off or compact support
deliver the same qualitative results.

One additional feature in our calculation in relation to
\cite{Zenginoglu:2008uc} is the arbitrary coordinate location of future null
infinity. This is accomplished by the following choice of the conformal
factor
\be\label{conf_fac} \Omega = 1-\frac{\rho}{S}. \ee
Here, $S$ denotes the coordinate location of $\scri^+$. Then the inverse of the coordinate compactification (\ref{eq:rcomp}) reads
\[ \rho = \frac{S r}{S+r}.\]
The event horizon is located at
\[ \rho_{EH} = \frac{2 m S}{2m+S}.\]
For the CMC foliation determined by (\ref{height_cmc}) there is a
minimal surface inside the event horizon where the coordinates break
down \cite{Malec:2003dq, Buchman:2009ew}. We avoid this coordinate
singularity in our excision technique by locating the inner boundary
of our computational domain inside and very close to the event
horizon. The outer boundary is located at $\rho = S$.

\begin{figure}[ht]
  \includegraphics[height=0.17\textheight,width=0.43\textwidth]{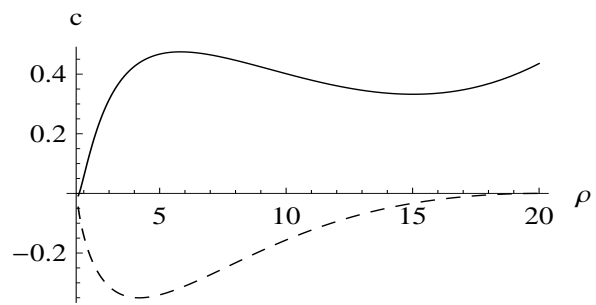}
  \caption{The dashed and solid curves represent coordinate speeds of
    in- and outgoing characteristics respectively for parameters $m=1, C=1, K=0.07, S=20$. There are no incoming characteristics into the computational
    domain. \label{fig:char_sp}}
\end{figure}

The in- and outgoing characteristic speeds for the system (\ref{comp_rwz}) are
given by 
\[ c_{\pm} = \frac{1}{L} (-\bbeta \pm\balpha^2).\]
Inserting the metric functions of a CMC foliation (\ref{eq:resc}) with
the conformal factor (\ref{conf_fac}) we get
\[ c_{\pm} = \bJ \sqrt{\bJ^2+ F\,\Omega^2} \pm (\bJ^2+ F\,\Omega^2).\]
Along null infinity we have
\be \label{eq:sp} c_+|_{\scri^+} = \frac{2}{9} K^2 S^2 , \qquad c_-|_{\scri^+} = 0.\ee
The outgoing characteristic speed at null infinity suggests that we
need to choose a smaller $K$ if we want to calculate gravitational
perturbations on a larger coordinate domain with an explicit time
integration scheme subject to the Courant
condition.  Equation (\ref{eq:sp}) suggests that $K$ and $S$ should satisfy $K\sim 1/S$. For the CMC calculations we set $m=1, C=1, K=0.07, S=20$. Fig.~\ref{fig:char_sp} shows the coordinate speeds of characteristics for these values of parameters. We see that there are no incoming characteristics into the simulation domain, hence no boundary conditions are needed. Outgoing radiation leaves the spacetime through the outer boundary.

\subsection{Quasinormal modes and tail decay rates}\label{sec:decay}
After an initial transient phase that depends on initial data, gravitational perturbations go through a quasinormal ringing phase followed by polynomial decay. In the following, we present results for the simulation of odd parity (Regge-Wheeler) perturbations with the potential (\ref{eq:odd_pot}). Simulations of even parity (Zerilli) perturbations deliver similar results.

\begin{figure}[ht]
  \includegraphics[width=0.37\textwidth]{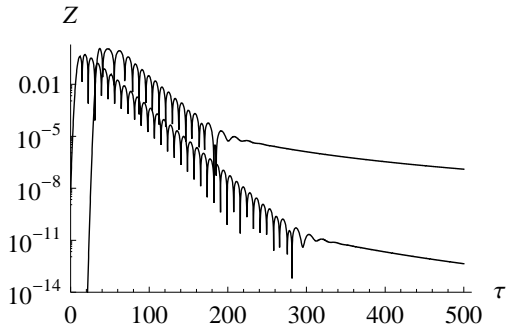}
  \caption{Quasinormal mode ringing of odd parity gravitational perturbations as measured along $\scri^+$ (top curve) and along an observer at $r=3m$ (bottom curve). \label{fig:qnm}}
\end{figure}

Fig.~\ref{fig:qnm} shows the quasinormal ringing and subsequent polynomial decay in a half-logarithmic plot for an observer at $r=3m$ and at null infinity. The ringing lasts longer for observers closer to the black hole due to the weaker tail signal. It has the form
\be \label{eq:qnm} Z(\tau) = a \,e^{-\omega_2 \tau} \sin(\omega_1 \tau + \varphi).\ee Here, $\omega_1$ and $\omega_2$
are the mode frequencies, $a$ is the amplitude and $\varphi$ is
the phase. We fit the wave signal along $r=3m$ to the above formula using a simple least squares method on the interval $\tau\in[60m,120m]$. We find
$\omega_1 = 0.373664$ and $\omega_2=0.088952$. These numerical values
are very close to those obtained by Leaver's continued fraction
method \cite{Leaver85, Leaver:1986gd}, which read $\omega_1 =0.373672 $ and $\omega_2=0.088963$. The error in the measured frequencies is dominated by fitting accuracy rather than by numerical accuracy, similarly as in \cite{Zenginoglu:2008uc}.

\begin{figure}[ht]
\centering
\includegraphics[width=0.37\textwidth]{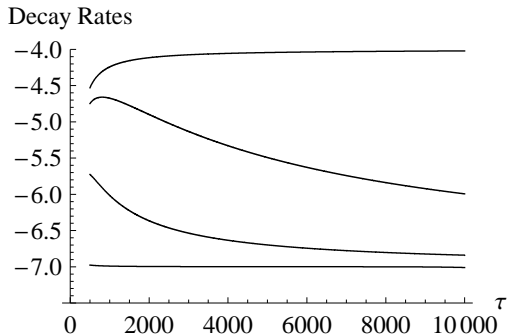}
\caption{The local power index for vacuum perturbations of a Schwarzschild black hole.  The locations of the observers are from top to bottom: null infinity, $2000m, 300m$ and $3m$. We used 2000 grid points and quadruple precision for the simulation. \label{fig:decay}}
\end{figure}

From the point of view of asymptotics, the tail part is more interesting than quasinormal mode ringing, because decay rates at null infinity and at
finite surfaces are different \cite{Gundlach:1993tp}. In Fig.~\ref{fig:decay} we plot, for various observers, the local power index defined as 
\be\label{eq:exponent} p_\rho(\tau) = \frac{d\ln |\phi(\tau,\rho)|}{d\ln \tau}, \ee 
for perturbations with $l=2$. The function $p_\rho(\tau)$ becomes asymptotically in time the exponent of the polynomial decay of the solution. 

Gravitational metric perturbations decay with rate $-(l+2)$ along null infinity and $-(2l+3)$ at finite distances from the black hole, where $l$ is the angular momentum number of the perturbation. The dominant $l=2$ mode is expected to decay as $-4$ at null infinity and as $-7$ at finite distances. This expectation is confirmed in Fig.~\ref{fig:decay} to a great accuracy. 

There is strong evidence that the asymptotic tail decay rates for
gravitational perturbations in the Newman-Penrose formalism at null
infinity and at finite distances are $-6$ and $-7$ respectively
\cite{Campanelli:2000in, Zenginoglu:2008uc}. The decay rate of metric perturbations agrees with the decay rate of the Weyl scalar $\psi_4$ near the black hole whereas the corresponding rates at null infinity differ. This behavior is explained by considering the relation between $\psi_4$ and metric perturbations. An asymptotic formula that one uses frequently in radiation extraction algorithms relates $\psi_4$ to second time derivatives of metric petrutbations. This relation explains the decay rates $-6$ for $\psi_4$ and $-4$ for $Z$ at null infinity. However, the full Chandrasekhar transformation relating $\psi_4$ to metric perturbations includes lower order terms that vanish asymptotically \cite{Chandrasekhar:1975, Lousto:2005xu}. These terms dominate near the black hole and therefore the rates for $\psi_4$ and $Z$ are the same in that region. Note that without having access to null infinity, the agreement between the finite distance decay rates for $\psi_4$ and $Z$ might be misleading. This observation supports the importance of including the asymptotic domain in numerical calculations.

\subsection{Matching}\label{sec:match}
Most applications of the RWZ formalism in modern literature include
the treatment of matter terms, such as a point particle. Elaborate algorithms are constructed to deal with phenomena like discontinuities and shocks related to modelling of matter. Many of these algorithms depend on a specific coordinate system typically based on Cauchy foliations. In this section we show how to use such a coordinate system near the black hole while using the hyperboloidal method in an exterior domain. 

The basic idea is to match a truncated Cauchy surface near the black hole to a hyperboloidal surface in the exterior region \cite{Zenginoglu:2007jw}. The matching is performed by a suitable choice of the height function derivative that determines the time transformation (\ref{time_traf}). We set
\be \label{eq:match_H} H = \left\{
\begin{array}{ll} \Hint,
  & \rho\leq \rhoi,  \\ 
    \Hint(1-f) + f \Hext, \qquad
   &  \rhoi<\rho<\rhoo,  \\ 
   \Hext, & \rho\geq \rhoo.
\end{array}\right.\ee 
Here, $\Hint$ describes a Cauchy foliation, $\Hext$ describes a hyperboloidal foliation and $f$ is a transition function that varies between $0$ and $1$. We set
\be \label{eq:f} f = \left\{
\begin{array}{ll} 0,
  & \rho\leq \rhoi,  \\ 
    f_T, \quad 
   &  \rhoi<\rho<\rhoo,  \\ 
   1, & \rho\geq \rhoo.
\end{array}\right.\ee 
In \cite{Zenginoglu:2007jw, Zenginoglu:2009hd} we used 
\be \label{eq:badf_t} f_T = 1-e^{-(\rho-\rhoi)^2/(\rho-\rhoo)^2}.\ee 
With the above choice of $f_T$, the function $f$ is smooth at $\rho=\rhoo$ but not at $\rho=\rhoi$. An everwhere smooth transition function has been suggested in \cite{Yunes:2005nn, Vega:2009qb}
\be \label{eq:f_t} f_T = \frac{1}{2}+\frac{1}{2}\tanh \left[
\frac{s}{\pi} \left( \tan\left(\frac{\pi}{2} \frac{\rho-\rhoi}{\rhoo-\rhoi}\right) - 
\frac{q^2}{\tan\left(\frac{\pi}{2} \frac{\rho-\rhoi}{\rhoo-\rhoi}\right) } \right) \right]. \ee
The free parameter $q$ determines the point $\rho_{1/2}$ at which $f_T(\rho_{1/2})=1/2$ and $s$ determines the slope of $f_T$ at $\rho_{1/2}$  \cite{Yunes:2005nn, Vega:2009qb}.

On the resulting matched surface we introduce a radial coordinate that coincides with the standard coordinate near the black hole and compactifies the exterior region. We set
\be \label{eq:match_Om} \Omega = \left\{
\begin{array}{ll} 1,
  & \rho\leq \rhoi,  \\ 
    (1-f) + f \left(1-\frac{\rho}{S}\right), \qquad
   &  \rhoi<\rho<\rhoo,  \\ 
   1-\frac{\rho}{S}, & \rho\geq \rhoo.
\end{array}\right.\ee 

\begin{figure}[ht]
  \centering
  \includegraphics[width=0.37\textwidth]{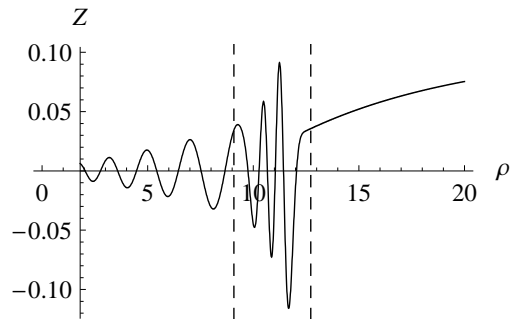}
  \caption{A snapshot from a matched surface evolution. On the left of the dashed lines we use an ingoing Eddington-Finkelstein foliation while on the right we use a CMC foliation. The dashed lines denote the area of transition. This plot visually demonstrates the spatial redshift property of hyperboloidal evolutions. \label{fig:snapshot}}
\end{figure}

As an example, we perform a matched surface evolution with an ingoing Eddington-Finkelstein foliation in the interior and a CMC foliation in the exterior.  We set $m=1, C=1.5, K=0.4, S=20, l=10$ and use 2000 radial grid points. Our time step is very small, $\Delta t=0.001$, due to the large value of $K$. The reason why we choose such high values for $l$ and $K$ is to demonstrate the spatial redshift effect along hyperboloidal surfaces discussed below. The radial domain is $\rho\in [1.8,20]$ implying that the inner radius is an excision surface and the outer radius is at null infinity. The inner and outer transition points are $\rhoi = 9.08$ and $\rhoo=12.72$. The transition region takes $20\%$ of the radial grid domain, the rest of the domain is shared equally by the ingoing Eddington-Finkelstein and the CMC foliations. We use the transition function (\ref{eq:f_t}) with $q=1.3$ and $s=1.5$. Fig.~\ref{fig:snapshot} shows a snapshot from this evolution taken at time $\tau=37m$. 

The signal along the Cauchy surface varies strongly over the grid, whereas the signal in the hyperboloidal part has a low frequency due to the redshift effect (Fig.~\ref{fig:snapshot}). This effect implies that hyperboloidal surfaces are particularly suitable for the accurate calculation of signals with strong spatial variation, and also that a small number of points is sufficient to resolve the signal which makes high order numerical methods very efficient in combination with the hyperboloidal approach. The spatial redshift effect is stronger for larger values of the mean curvature $K$ or higher speeds of outgoing characteristics. 

\begin{figure}[ht]
  \centering
  \includegraphics[width=0.37\textwidth]{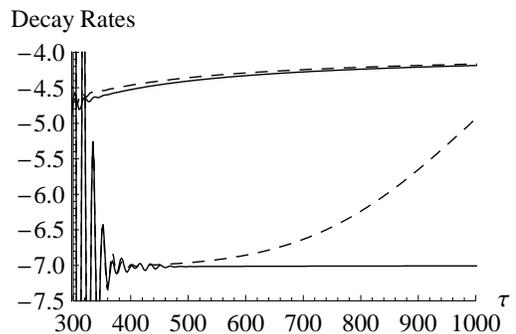}
  \caption{The decay rates at null infinity (top curve) and at $r=3m$ (bottom curve). For the dashed curves we use the transition function (\ref{eq:badf_t}), for the solid curves we use (\ref{eq:f_t}) with the same numerical setting otherwise. The smooth transition function (\ref{eq:f_t}) leads to a more accurate calculation than (\ref{eq:badf_t}).  \label{fig:transit_rates}}
\end{figure}

The transition function (\ref{eq:f_t}) leads to a more accurate evolution than (\ref{eq:badf_t}). To demonstrate the difference, we plot in Fig.~\ref{fig:transit_rates} the tail decay rates for an $l=2$ perturbation calculated with the two transition functions. The dashed line corresponds to the decay rates as measured from an evolution with (\ref{eq:badf_t}) whereas the solid line is calculated from an evolution with (\ref{eq:f_t}) using the same parameters otherwise. We see that, for the set of parameters we used, the decay near the black hole cannot be captured accurately by the evolution with the transition function (\ref{eq:badf_t}) whereas the evolution with (\ref{eq:f_t}) is accurate.

\section{Discussion} \label{sec:diss}
We performed a hyperboloidal study of gravitational
perturbations of Schwarzschild spacetime using the
Regge-Wheeler-Zerilli formalism extending an earlier work based on the
Bardeen-Press formalism \cite{Zenginoglu:2008uc}. The motivation for
the current work comes from the simplicity and the wide application of
the RWZ formalism in studies of phenomena related to gravitational radiation.

With the hyperboloidal method we can calculate in a simple and efficient way gravitational perturbations of Schwarzschild spacetime in time domain using the
common RWZ formalism without introducing an artificial outer boundary
into the spacetime. On the physical side, we gain some insight into the
validity of asymptotic formulae on gravitational radiation. Studies of decay rates of
gravitational perturbations at finite distances may lead to the
erroneous conclusion that the Newman-Penrose scalar $\psi_4$ decays
with the same rate as metric perturbations. While the rates are the same at finite distances, they are different at null infinity in accordance
with the asymptotic formula relating $\psi_4$ to second time derivatives of metric
perturbations. This observation emphasizes the importance of including the asymptotic domain in numerical calculations of gravitational radiation.

On the hyperboloidal side, our study employs an improved
coordinatization of Schwarzschild spacetime that may be useful in the
application of our method to problems including matter terms near the
black hole. The generality in coordinatization is important because many algorithms handling matter terms rely on jump conditions or regularization procedures that have been implemented in certain coordinate systems. In Sec.~\ref{sec:match} we combined horizon penetrating coordinates in the interior (\ref{height_ef}) with a CMC foliation in the exterior (\ref{height_cmc}) but other choices are possible. One can, for example, use Schwarzschild time slices with tortoise coordinates in the interior and slices based on the asymptotic hyperboloidal condition (\ref{height_taylor}) in the exterior. The implementation of the hyperboloidal method in applications of the RWZ formalism should therefore require only minor modifications. In our study of the matching, the smooth transition function (\ref{eq:f_t}) applied in \cite{Yunes:2005nn, Vega:2009qb} turned out to be superior to the simple transition function (\ref{eq:badf_t}) used in \cite{Zenginoglu:2007jw, Zenginoglu:2009hd}. We note, however, that the implementation of the matching is an additional source of error. Considering that wave extraction methods are rather reliable when the background has been given, the hyperboloidal method may be less efficient than the implementation of accurate boundary conditions such as those presented in \cite{Lau:2004as, Lau:2004jn, Lau:2005ti}. The choice of method needs to be made on a case by case basis depending on the accuracy requirements, time scale of the simulation and which features of the radiation signal are of interest.

A natural next step in the hyperboloidal approach for the RWZ formalism is to handle source terms near the black hole. Among further applications of the method are Cauchy-perturbative matching with hyperboloidal surfaces in the exterior \cite{Rupright98, Rezzolla99a, Zink:2005pe}, or the study of second order perturbations of Schwarzschild spacetime at null infinity \cite{Brizuela:2009qd, Brizuela:2006ne}. Perturbative results may not be valid even in certain weak field situations \cite{Bizon:2007qz,Bizon:2007xa, Bizon:2008iz}. It is therefore important to check such results by performing comparisons with nonlinear evolution.

Eventually, the main problem one would like to tackle with scri fixing is the hyperboloidal initial value problem for the Einstein equations \cite{Zenginoglu:2008pw, Moncrief:2008ie}. Recently, Rinne presented the first successful numerical implementation of hyperboloidal scri fixing for the Einstein equations in axial symmetry \cite{Rinne:2009qx}. It would be interesting to check the validity of linear perturbation theory at null infinity by comparing results obtained in this paper with Rinne's fully nonlinear calculations. 

\acknowledgments This work was supported in part by the NSF grant
PHY0801213 to the University of Maryland. I thank Peter Diener for drawing my attention to the smooth transition function (\ref{eq:f_t}). I thank Luisa Buchman, Peter Diener, Sascha Husa, Stephen Lau, Vincent Moncrief, Dar{\'i}o N\'u\~nez, Oliver Rinne and Manuel Tiglio for discussions and comments on the manuscript.

\bibliography{references} \bibliographystyle{h-physrev5}

\end{document}